\documentclass[conference]{IEEEtran}
\IEEEoverridecommandlockouts
\usepackage{amsmath,amssymb,amsfonts}
\usepackage{algorithmic}
\usepackage{graphicx}
\usepackage{textcomp}
\usepackage{xcolor}
\usepackage{subcaption}
\usepackage{microtype}
\usepackage{multicol}
\usepackage{bm}
\usepackage{booktabs}
\usepackage{cite}
\def\BibTeX{{\rm B\kern-.05em{\sc i\kern-.025em b}\kern-.08em
    T\kern-.1667em\lower.7ex\hbox{E}\kern-.125emX}}
\begin{document}

\title{Sorted Weight Sectioning for Energy-Efficient Unstructured Sparse DNNs on Compute-in-Memory Crossbars
}

\author{\IEEEauthorblockN{Matheus Farias}
\IEEEauthorblockA{\textit{Harvard University} \\
matheusfarias@g.harvard.edu}
\and
\IEEEauthorblockN{H. T. Kung}
\IEEEauthorblockA{
\textit{Harvard University}\\
kung@harvard.edu}
}

\maketitle

\begin{abstract}
We introduce \textit{sorted weight sectioning} (SWS): a weight allocation algorithm that places sorted deep neural network (DNN) weight sections on bit-sliced compute-in-memory (CIM) crossbars to reduce analog-to-digital converter (ADC) energy consumption. Data conversions are the most energy-intensive process in crossbar operation. SWS effectively reduces this cost leveraging (1) small weights and (2) zero weights (weight sparsity).

DNN weights follow bell-shaped distributions, with most weights near zero. Using SWS, we only need low-order crossbar columns for sections with low-magnitude weights. This reduces the quantity and resolution of ADCs used, exponentially decreasing ADC energy costs without significantly degrading DNN accuracy. 

Unstructured sparsification further sharpens the weight distribution with small accuracy loss. However, it presents challenges in hardware tracking of zeros: we cannot switch zero rows to other layer weights in unsorted crossbars without index matching. SWS efficiently addresses unstructured sparse models using offline remapping of zeros into earlier sections, which reveals full sparsity potential and maximizes energy efficiency.

Our method reduces ADC energy use by 89.5\% on unstructured sparse BERT models. Overall, this paper introduces a novel algorithm to allow energy-efficient CIM crossbars for unstructured sparse DNN workloads.
\end{abstract}


\section{Introduction}
Resistive compute-in-memory (CIM) crossbars have emerged as promising deep neural network (DNN) accelerators thanks to their ability to mitigate costly data movements between memory and processing units \cite{horowitz2014, Huang_2020,fouda2022}. By integrating storage and computation directly within the memory array, crossbars perform multiply-accumulate (MAC) operations with higher speed and lower power consumption compared to conventional digital systems \cite{chakraborty2020, shafiee2016, donhee2022, abu2020, huo2022}. However, the power efficiency of CIM architectures is limited by the energy consumption of analog-to-digital converters (ADCs), which can account for up to 85\% of the total energy and area in these systems \cite{li2015}. This presents a critical challenge to the practical deployment of CIM-based accelerators for DNN workloads.

One strategy to enhance the energy efficiency of CIM crossbars is to exploit sparsity in DNNs. Sparsity, achieved through various pruning techniques, reduces the number of weights, thereby lowering required memory and MAC operations. Pruning can be either unstructured and structured. While unstructured pruning eliminates individual weights to increase sparsity with small model accuracy drop, it hardens efficient tracking due to irregular patterns of zero weights.

On the other hand, structured pruning removes entire neurons, filters, or channels, creating regular patterns of zeros which eases hardware tracking. This method reduces crossbar mapping complexity, leading to better energy efficiency. However, structured pruning can result in significant model accuracy loss due to the aggressive reduction in the network’s capacity to learn and represent complex features \cite{blalock2020state}. Thus, combining the benefits of unstructured pruning’s reduced accuracy drop with the hardware-friendly properties of structured pruning is a key challenge to balance efficiency and accuracy.

To address this challenge, we propose a novel technique called \textit{sorted weight sectioning} (SWS), which optimizes unstructured sparse DNN implementation on CIM crossbars. SWS strategically organizes weights exploiting two critical DNN characteristics to minimize ADC energy consumption while maintaining model accuracy: (1) bell-shaped distribution, where a large proportion of weights are distributed around zero, and (2) weight sparsity. By sorting weights by magnitude and grouping them into sections, SWS maps smaller weights to crossbar columns with lower power-of-two multipliers, termed \textit{low-order columns}. These sections require less precise and fewer ADCs, thereby reducing energy use without compromising model accuracy. Also, the technique remaps isolated zero weights, achieving both energy efficiency and robust model performance.

\begin{figure*}
    \centering
    \includegraphics[width = 0.7\textwidth]{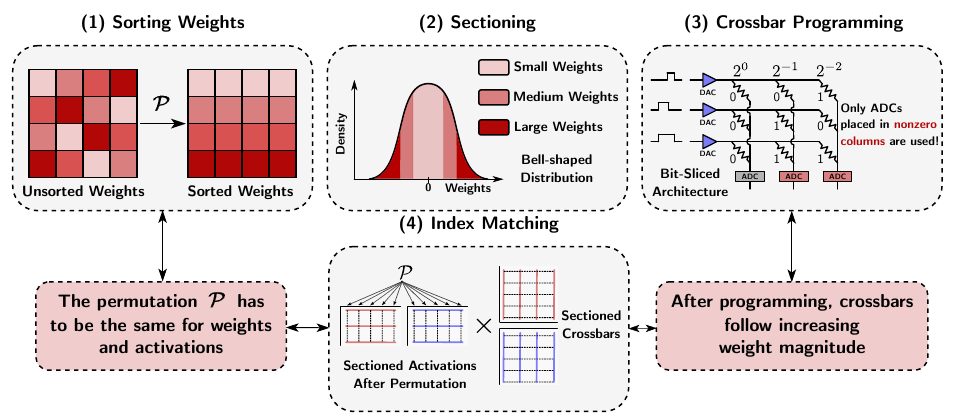}
    \caption{Summary of the approach. (1) sort the weights, (2) partition into sections of increasing weight magnitude, (3) program each crossbar section, and (4) permute the activation vector to ensure dot-product correctness.}
    \label{fig:summary}
\end{figure*}

SWS involves four steps: (1) sorting the weight vector by magnitude, (2) partitioning the sorted vector into sections of rows, where each row corresponds to a weight value, (3) programming each section into a CIM crossbar, and (4) permuting the activation vector according to the sorted weight order to ensure dot-product correctness. This approach, summarized in Figure \ref{fig:summary}, enables fine-grained control of ADC usage and enhances energy efficiency of CIM architectures for sparse DNNs.
The main contributions of this paper are:
\begin{itemize}
    \item A mathematical analysis of how programming sections with sorted weights reduce the ADC energy consumption.
    \item A novel permutation strategy based on sorted sectioning leveraging DNN weight distribution and sparsity to decrease ADCs energy use without compromising accuracy.
    \item End-to-end experiments in the state-of-the-art simulator CiMLoop\cite{cimloop} and PyTorch to show energy consumption reduction after applying sorted weight sectioning.
\end{itemize}

\section{Background}
\label{sec:back}

In resistive crossbars, weights are stored in the conductance of programmable resistors called memristors \cite{Wang2015, liao2018} and grouped in 1T1R cells to simplify programming \cite{zangeneh2014}. This work uses bit-sliced design where each crossbar row represents one bit-sliced weight value and each column is a negative power-of-two multiplier \cite{shafiee2016, chou2019}.
Bit-sliced architectures are commonly used for precision-demanding applications \cite{abu2020}.
Despite using more memristors than multi-level implementations \cite{huo2022, Cai2019}, bit-sliced architectures fit signed weights without additional digital logic and with fewer nonidealities \cite{chakraborty2020}.

The dot product is computed by accumulating currents obtained by multiplying electric potentials with conductances across each column.
Finally, ADCs convert currents back to digital.
Digitized column outputs are shifted according to their column index (i.e., multiplied by a power of two). Past works address digital partial sum accumulation to increase accuracy \cite{shafiee2016, sun2018, zidan2013}.
Small crossbars reduce nonidealities but require more ADCs: one for each partial sum accumulation section.
Combining sectioning with bit-slicing is discussed in \cite{chou2019}.

ADCs are the main energy bottleneck, consuming up to 85\% of the area and energy for mixed-signal tasks \cite{li2015}.
A $b$-bit Flash ADC requires $2^b$ active comparators for conversion. Moreover, one of the most common ADC architectures, the SAR-ADC, has a built-in DAC module.
Thus, we target analog-to-digital conversions to minimize crossbar energy consumption. 




\begin{figure}
    \centering
    \includegraphics[width = \columnwidth]{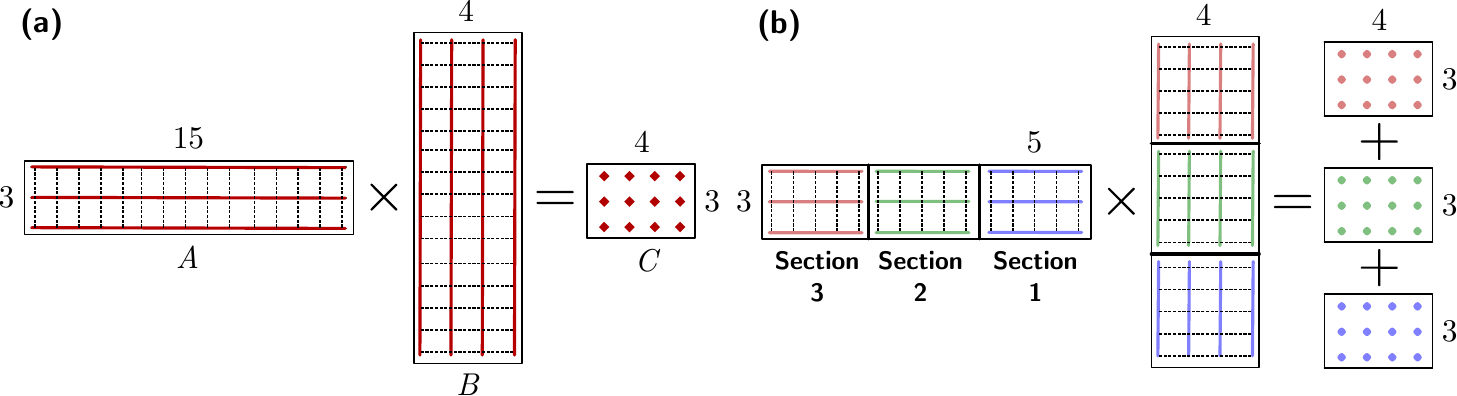}
    \caption{(a) The dot product of a $15$-element row vector (red) of $A$ and the $15$-element column vector (red) of $B$ to compute final elements of $C$, and (b) the dot product of a colored section, such as the section 1 (blue) of $A$ and the corresponding section (blue) of $B$ computes partial results for $C$.
    Accumulating dot products for partitioned $A$ and $B$ in Figure \ref{fig:sectioning}b generalize conventional dot products in Figure \ref{fig:sectioning}a by using matrix sections as matrix elements.}
    \label{fig:sectioning}
\end{figure}

\begin{figure}
    \centering
    \includegraphics[width =\columnwidth]{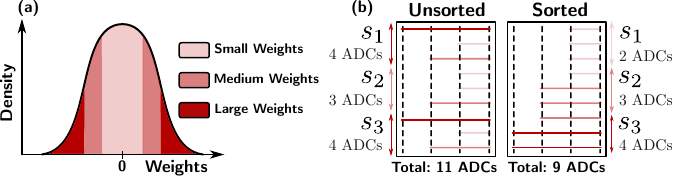}
    \caption{
    (a) Pretrained weights follow a bell-shaped distribution around zero. (b)
    This example compares the number of ADCs required for sections $\bm{s_1}$, $\bm{s_2}$, and $\bm{s_3}$ using unsorted and sorted approaches. As most frequent weights are near zero, we require only low-order columns, thereby, using fewer ADCs.
    }
    \label{fig:unsorted_sorted}
\end{figure}

\section{Sorted Weight Sectioning}
\label{sec:sorted_sec}

In crossbar computing, dividing the matrix-matrix multiplication (MMM) into sections generalizes the conventional approach by segmenting the operation based on the architecture's constraints, such as size or precision limits. Each section corresponds to a subset of the matrix, and the results are aggregated to compute the final output (see Figure \ref{fig:sectioning}). This sectioning not only mitigates hardware limitations but also opens the door to novel optimizations.

\subsection{Mathematical Formulation of SWS}
Now we justify why SWS provides a weight allocation strategy that enhances energy efficiency.

Let $W$ be a random variable following a normal distribution $N(0, \sigma)$ (approximation of bell-shaped curve of pretrained weights). Define the symmetric regions:
\begin{equation}
    \begin{split}
        S_k &= (-w_k, -w_{k-1}) \cup (w_{k-1}, w_k) \quad \text{and} \\
        S_{k+1} &= (-w_{k+1}, -w_k) \cup (w_k, w_{k+1}),
    \end{split}
\end{equation}
where $0 < w_{k-1} < w_k < w_{k+1}$ (sorted weight assumption).

Suppose $w \in S_k$ and $w' \in S_{k+1}$ are represented in binary form up to $b$ bits:
\begin{equation}
w = \sum_{i=0}^{b-1}a_{i}2^{-i} \quad \text{and} \quad w' = \sum_{i=0}^{b-1}a'_{i}2^{-i},    
\end{equation}
where $a_i, a'_i \in \{0, 1\}$ for $ i = 0, 1, \dots, b - 1$.

\textbf{Goal:} Show that for any $i$ with $0 \leq i \leq b - 1$:
\begin{equation}
P(a_i = 0) > P(a'_i = 0),    
\end{equation}
where $P(a_i = 0)$ is the probability of $a_i=0$. In crossbars, if $a_i = 0$, it means the weight will not contribute to placing an ADC at the column end.

Intuitively, because $S_k$ is closer to zero than $S_{k+1}$, and the normal distribution is symmetric and decreasing as $|w|$ increases, the probability density $f(w) = \frac{1}{\sqrt{2\pi\sigma^2}} e^{-w^2/2\sigma^2}$ is higher and decreases more steeply in $S_k$. This causes a greater imbalance between the probabilities of $a_n = 0$ and $a_n = 1$ in $S_k$ compared to $S_{k+1}$. As a result, the chance that $a_n = 0$ is greater for $w$ in $S_k$ than for $w'$ in $S_{k+1}$.

Mathematically, For each bit position $n$, the bits $a_0,\dots, a_{n-1}$ define an interval $[L, U]$ for $|w|$:
\begin{equation}
L = \sum_{i=0}^{n-1} a_i 2^{-i}, \quad U = L + 2^{-n}.
\end{equation}
The bit $a_n$ divides this interval into two subintervals:
\begin{itemize}
    \item If $ a_n = 0 $, $|w| \in [L, M]$ 
    \item If $a_n = 1$, $|w| \in [M, U]$
\end{itemize}
where $M = \frac{L + U}{2}$. Equivalently, for $[L', U']$, the midpoint $M'$.

Since $\frac{\int_L^M f(w) \, \text{d}w}{\int_L^U f(w) \, \text{d}w}$ monotonically decreases with $L$, the probabilities follow:
\begin{equation}
    P(a_n = 0) = \frac{\int_L^M f(w) \, \text{d}w}{\int_L^U f(w) \, \text{d}w} > \frac{\int_{L'}^{M'} f(w) \, \text{d}w}{\int_{L'}^{U'} f(w) \, \text{d}w} = P(a'_n = 0).
\end{equation}
Therefore, earlier sections require fewer ADCs.
\subsection{Qualitative Analysis of SWS}
In the previous subsection, we reduced the ADC count for a fixed sectioning size using \textit{sorted weight sectioning} (SWS). The method consists of two offline steps. We first sort crossbar rows by their magnitudes, defining a section-specific index matching function (see $\mathcal{P}$ in Figure \ref{fig:summary}). Then, we section the MMM so sections are ordered in increasing magnitude.

Low-magnitude weight sections only activate low-order columns and the ADC count is determined by columns that accumulated non-zero values (see Figure \ref{fig:unsorted_sorted}).
Thus, earlier sections are filled with zeroed-entry columns which do not need data conversions. Consequently, these sections need fewer ADCs. Note that zero activations are not fetched into crossbars because they output zeros in all columns. We also do not program zero weights since they are rows filled with zeros.

This approach has a side effect of remapping isolated zeros in unstructured sparse models to early sections, which makes a perfect use case for SWS. We leverage this hardware-friendly sparsity mapping while having small accuracy drop compared to structured sparsity. With SWS, sparsity reduces the number of sections since sections full of zeros are not programmed.

We reuse the input permutation function on-the-fly for every inference task, ensuring index matching. This process' memory/latency depends on the feature size since we must allocate memory to hold and permute the fetched data while maintaining constant computation throughput. 

To minimize permutation latency, we use $f$ muxes for an input with feature size $f$. The mux selector is the pre-computed new index for each element. This implementation performs full permutation in one clock cycle. For $L$ crossbars, the space complexity of the approach is $\mathcal{O}(f)$ and the time complexity is $\mathcal{O}(fL\log{f})$. To mitigate sorting overhead, we reuse the data buffer to save memory and crossbars to reduce latency.

\subsection{Impact on Energy and Accuracy}
\label{sec:impact}
The impact of SWS on model accuracy and energy consumption depends on the DNN weight distribution.
DNN weights follow a bell-shaped distribution with long tails \cite{han2015, fang2020, horton2022, tambe2020} due to normalization layers during training.
This ensures more sections of small weights than of large.

High-order columns are filled with zeros in sections with small magnitude weights, which makes these sections require fewer ADCs.
Furthermore, low-order column outputs are scaled by smaller power-of-two multipliers, motivating the use of different ADC resolutions per column.
We can use lower precision conversions for columns farther from the crossbar.
As a result, we can use both a smaller number of ADCs and a lower resolution to reduce energy costs without significantly degrading DNN accuracy.
We note that energy savings are more pronounced for models with sharper weight distributions around zero and longer tails -- models with smaller $\sigma$ will require fewer ADCs due to the higher occurrence of zero bits.

\section{Experiments}
\label{sec:exp}
We simulate crossbar computation in PyTorch and CiMLoop on ImageNet-1K \cite{deng2009}, CIFAR-10 \cite{kritz2009}, and MNIST \cite{deng2012} datasets. We apply our method on all layers of models from PyTorch (ResNets and VGGs), \texttt{timm} (ViTs and DeITs patch 16 224), and \texttt{transformers} (BERTs) libraries trained in 32-bit floating point.
In all cases, crossbars have 8 columns.
We consider two scenarios for evaluation:
\begin{itemize}
         \item \textbf{Sorted (our approach)}: crossbar rows are sorted by their magnitudes. This scenario can be with fixed resolution (ADCs with the same resolution) or unfixed resolution.
     \item \textbf{Unsorted (state-of-the-art approach considered in \cite{shafiee2016, chou2019})}: crossbar rows are placed randomly.
\end{itemize}

\noindent

\subsection{Results}
\label{sec:sortedpermutations}
\begin{figure}
    \centering
    \includegraphics[width = \columnwidth]{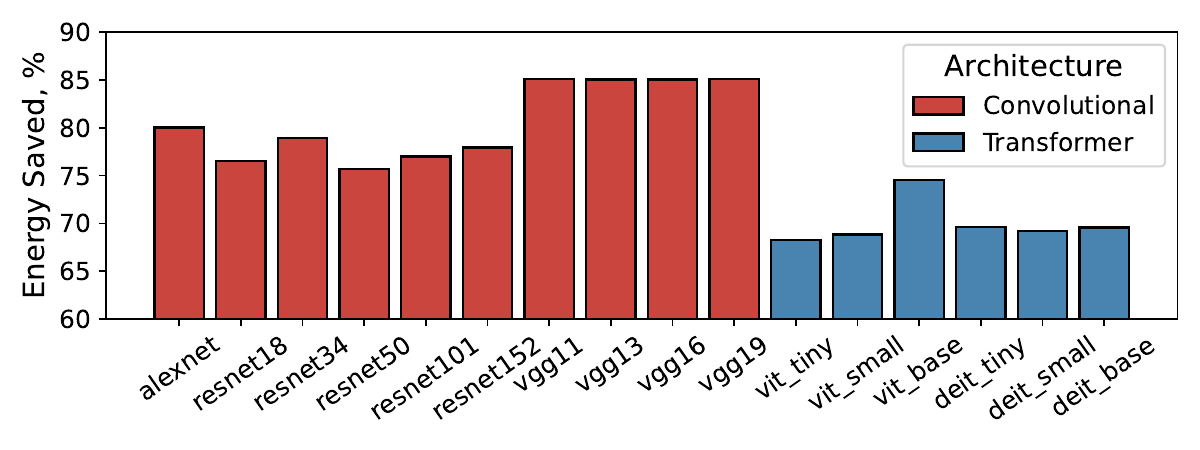}
    \caption{
    ADC energy savings applying SWS with fixed ADC resolution on convolutional (red) and vision transformer (blue) models for ImageNet-1K.
    }
    \label{fig:wres}
\end{figure}

We fix ADC resolution to $10$ bits and use 128-row sectioning for many DNN models on ImageNet-1K.
We decreased the number of ADCs used thereby the energy consumption with SWS (see Figure \ref{fig:wres}). 
Because we only apply SWS fixing ADC resolutions at this point, we did not perceive a significant accuracy drop. In fact, the largest drop was 0.09\% in DeIT-Tiny. 
SWS achieves better energy savings for CNNs because transformer models have (1) wider weight distributions (noted in Section \ref{sec:sorted_sec}) and (2) relevant outliers, which makes quantization harder \cite{zeroquant,han2023, quant2023,tambe2020}. Furthermore, we used CIMLoop to get energy values and account for the sorting hardware driver (see Table \ref{tab:cimloop}). We highlight CIM Unit and ADC energy consumption reduction. This happens with SWS because sparsity reduces the number of sections and skip ADCs.
\begin{table}[]
\centering
\caption{Energy breakdown for language models in CIMLoop.}
\resizebox{\columnwidth}{!}{%
\begin{tabular}{cccccl}
\toprule
\textbf{Baseline Models} & \textbf{Drivers + DAC (pJ/MAC)} & \textbf{CIM Unit (pJ/MAC)} & \textbf{ADC (pJ/MAC)} & \textbf{Total (pJ/MAC)} & \multicolumn{1}{c}{\textbf{Parameters}} \\ \midrule
MobileBERT & 1.169801 & 0.956911 & 14.1433 & 16.27 & \multicolumn{1}{c}{25M} \\
Microsoft Phi & 0.3078242 & 0.265546 & 3.9607 & 4.5340702 & \multicolumn{1}{c}{1.3B} \\
GPT-2 Medium & 0.346854 & 0.299687 & 2.4511 & 3.097641 & \multicolumn{1}{c}{355M} \\
\multicolumn{1}{l}{BERT 80\% Pruned} & 0.30253 & 0.02443 & 0.5132 & 0.84016 & \multicolumn{1}{c}{68M} \\
\multicolumn{1}{l}{BERT 85\% Pruned} & 0.32412 & 0.02171 & 0.4588 & 0.80463 & \multicolumn{1}{c}{51M} \\
\multicolumn{1}{l}{BERT 90\% Pruned} & 0.31555 & 0.01929 & 0.3343 & 0.66914 & \multicolumn{1}{c}{34M} \\ \hline
\textbf{Sorted Models} &  &  &  &  &  \\ \cline{1-1}
MobileBERT & 1.549332 ($\uparrow32.44\%$) & 0.935127 ($\downarrow 2.27\%$) & 3.3212 ($\downarrow 76.52\%$) & 5.805659 ($\downarrow 64.32\%$) & \multicolumn{1}{c}{25M} \\
Microsoft Phi & 0.4174333 ($\uparrow35.61\%$) & 0.245217 ($\downarrow 7.65\%$) & 0.9827 ($\downarrow 75.19\%$) & 1.64535 ($\downarrow 63.71\%$) & \multicolumn{1}{c}{1.3B} \\
GPT-2 Medium & 0.444351 ($\uparrow 28.11\%$) & 0.287684 ($\downarrow 4.00\%$) & 0.6313 ($\downarrow 74.24\%$) & 1.36335 ($\downarrow 55.99\%$) & \multicolumn{1}{c}{355M} \\
\multicolumn{1}{l}{BERT 80\% Pruned} & 0.36512 ($\uparrow 20.69\%$) & \textbf{0.01832 (}$\bm{\downarrow 25.01}$\textbf{\%)} & \textbf{0.0733 (}$\bm{\downarrow 85.72}$\textbf{\%)} & 0.45674 ($\downarrow 45.64\%$) & \multicolumn{1}{c}{68M} \\
\multicolumn{1}{l}{BERT 85\% Pruned} & 0.39834 ($\uparrow 22.90\%$) & \textbf{0.01593 (}$\bm{\downarrow 26.62}$\textbf{\%)} & \textbf{0.0550 (}$\bm{\downarrow 88.01}$\textbf{\%)} & 0.46927 ($\downarrow 41.68\%$) & \multicolumn{1}{c}{51M} \\
\multicolumn{1}{l}{BERT 90\% Pruned} & 0.38848 ($\uparrow 23.11\%$) & \textbf{0.01311 (}$\bm{\downarrow 32.04}$\textbf{\%)} & \textbf{0.0352} \textbf{(}$\bm{\downarrow 89.47}$\textbf{\%)} & 0.43679 ($\downarrow 34.72\%$) & \multicolumn{1}{c}{34M} \\ \bottomrule
\end{tabular}%
}
\label{tab:cimloop}
\end{table}
\begin{table}[]
\centering
\caption{Energy and accuracy of various sectioning methods.}
\resizebox{\columnwidth}{!}{%
\begin{tabular}{cccc}
\toprule
\textbf{ResNet-50 (ImageNet-1K)}                    & \textbf{Accuracy} & \textbf{Energy Consumption} & \textbf{ADCs}\\ \hline
Unsorted Sectioning                             & 78.10\%                & --  &[10-10-10-10-10-10-10-10]                       \\
Sorted Sectioning (Fixed Resolution)         & 78.03\% (\textdownarrow 0.07\%)                & \textdownarrow 75.70\%      &[10-10-10-10-10-10-10-10]                    \\
Sorted Sectioning (Unfixed Resolution) & 77.17\%  (\textdownarrow 0.92\%)              & \textdownarrow 81.64\%               &[10-10-10-10-10-9-9-8]            \\ \hline
\textbf{ViT-Base (ImageNet-1K)}                     &                   &       &                      \\ \hline
Unsorted Sectioning                              & 76.21\%                & --      &[10-10-10-10-10-10-10-10]                   \\
Sorted Sectioning (Fixed Resolution)        & 76.13\% (\textdownarrow 0.08\%)                & \textdownarrow 74.54\%             &[10-10-10-10-10-10-10-10]             \\
Sorted Sectioning (Unfixed Resolution) & 75.08\% (\textdownarrow 1.13\%)                & \textdownarrow 79.05\%                    &[10-10-10-10-10-10-9-8]      \\ \hline
\textbf{VGG-11 (CIFAR-10)}                    &  & & \\ \hline
Unsorted Sectioning                             & 88.78\%                & --    &[8-8-8-8-8-8-8-8]                   \\
Sorted Sectioning (Fixed Resolution)         & 88.71\% (\textdownarrow 0.07\%)                & \textdownarrow 73.46\%    &[8-8-8-8-8-8-8-8]                        \\
Sorted Sectioning (Unfixed Resolution) & 87.98\% (\textdownarrow 0.8\%)                & \textdownarrow 88.50\%           &[8-8-8-8-8-7-7-6]                \\ \hline
\textbf{LeNet-5 (MNIST)}                     &                   &                   &          \\ \hline
Unsorted Sectioning                            & 98.34\%               & --           &[6-6-6-6-6-6-6-6]              \\
Sorted Sectioning (Fixed Resolution)         & 98.37\% (\textuparrow 0.03\%)                & \textdownarrow 14.8\%        &[6-6-6-6-6-6-6-6]                 \\
Sorted Sectioning (Unfixed Resolution) & 98.36\% (\textuparrow 0.02\%)                & \textdownarrow 74.00\%                &[6-6-6-6-5-4-4-0]         \\ \bottomrule
\end{tabular}%
}
\label{tab:expsorting}
\end{table}

We further decreased energy consumption by decreasing ADC resolutions in each column as discussed in Section \ref{sec:sorted_sec} (see unfixed resolution results in Table \ref{tab:expsorting}). Table \ref{tab:expsorting} presents the ADC resolution from the highest to the lowest-order column, selected to balance energy savings and accuracy. We could specifically prune the lowest-order column ADC on LeNet-5 to boost energy efficiency without compromising accuracy.

Deeper networks are more likely to present better energy-saving results since the distribution of later layers is sharper. We observe this when comparing CIFAR-10 and MNIST results. ImageNet required higher-resolution ADCs. Although deeper networks are expected to have better energy-saving results, complex datasets are more sensitive to quantization. This justifies why small modifications on ADC resolutions already dropped 1.13\% of accuracy on ViT-Base.

\section{Discussion}
ISAAC\cite{shafiee2016} and CASCADE\cite{chou2019} used fixed ADC resolutions in their architectures.
Section \ref{sec:sortedpermutations} showed that DNNs have different active column distributions and may require less ADC resolution.
In our case, reconfigurable ADCs are attached to each column.
We can set the ADC precision for active columns, and disable when inactive.

Recent works propose weight allocation algorithms to increase crossbar energy efficiency.
McDanel et al. \cite{mcdanel2021} use term quantization to increase sparsity in crossbars, reducing ADC resolutions.
Han et al. \cite{han23} maximize data reuse on convolutional neural network (CNN) layers to reduce ADC use.
Huang et al. \cite{huang23} manage a hybrid in/near memory system allocating weights on each system based on their sensitivity.
Our work is orthogonal to these as it is (1) encoding-agnostic, (2) model-agnostic (supporting, e.g., CNNs and transformers), and (3) performed entirely within memory.

ISAAC's flipping encoding reduces ADC resolution by one bit.
However, it places sample-and-hold circuits on each crossbar column, significantly increasing latency. SWS reduces ADC resolution without considerable latency overhead in Section \ref{sec:sorted_sec}.
CASCADE uses pure analog accumulation with bit-sliced architecture. Sectioning is preferred due to finer-grained quantization and less accumulation of analog nonidealities.

For future work, we can model ADC sampling frequency similar to CASCADE, selecting lower frequencies for higher-order columns to further reduce energy use.
We can also consider nonidealities (e.g., sneak paths) when designing efficient weight allocations.

\section{Conclusion}
We showed how sorting and sectioning pretrained DNN weight for CIM crossbars yields significant savings on ADC energy consumption.
The approach's success lies in its ability to exploit bell-shaped DNN weight distributions (see Figure \ref{fig:unsorted_sorted}) and unstructured sparsity.
To the best of our knowledge, we are the first to observe this.

Our strategy works for any computation casted as matrix multiplication.
Its effectiveness depends on the application's tolerance to approximate computing and how values are distributed: the sharper the distribution, the fewer ADCs are used (and the lower the ADC resolution).
For larger models, we can increase (1) weight bitwidths by adding more columns to the crossbar and (2) output precision by increasing ADC resolution for each column output.
This enables flexible trading of accuracy for energy savings. SWS reduces ADC energy use by 89.5\% on pruned BERT models.

The idea is simple to implement as it merely involves proper weight placements on the crossbar.
We have provided end-to-end validation of these results via simulations for DNNs on well-known datasets in the literature.
This paper suggests a fruitful new direction for future research on CIM implementations for DNNs based on sorted weight sectioning.

\bibliographystyle{ieeetr}
\bibliography{refs}
\end{document}